\documentstyle[14lomcon,cite,epsfig,amssymb]{article}

\begin{document}

\title{NEUTRINOLESS DOUBLE BETA DECAY: SEARCHING FOR NEW PHYSICS WITH COMPARISON OF DIFFERENT NUCLEI}

\author{ A. Ali \footnote{e-mail: ahmed.ali@desy.de}}

\address{Deutsches Electronen-Synchrotron, DESY, 22607 Hamburg, Germany}

\author{ A. V. Borisov \footnote{e-mail: borisov@phys.msu.ru}}

\address{Faculty of Physics, Moscow State University, 119991 Moscow, Russia}

\author{ D. V. Zhuridov \footnote{e-mail: zhuridov@phys.nthu.edu.tw}}

\address{Department of Physics, National Tsing Hua University, 30013 Hsinchu, Taiwan}


\maketitle\abstracts{~The neutrinoless double beta decay is analyzed using a general Lorentz invariant
effective Lagrangian for various decaying nuclei of current experimental interest: $^{76}$Ge, $^{82}$Se, $^{100}$Mo,
$^{130}$Te, and $^{136}$Xe. We work out the half-lives and angular correlation coefficients of the outgoing electrons in several
scenarios for new physics: the left-right symmetric models, the R-parity-violating SUSY
and models with leptoquarks. The theoretical uncertainty in the nuclear matrix elements is discussed.}

~~~{\bf 1.} The Majorana nature of neutrino masses is anticipated by most of the
theories created to explain the observable lightness
of neutrinos, in particular, seesaw mechanism and
models with radiative neutrino mass generation~(see \cite{moh,DCH} and references therein).
Experimental evidence for the neutrinoless double beta decay
($0\nu2\beta$) would deliver a conclusive confirmation of the
Majorana nature of neutrinos, in contrast to the Dirac nature of all
other known fermions. This is the overriding interest in carrying
out these experiments and in the related
phenomenology~\cite{0nu2beta}. We recall that $0\nu2\beta$ decays
are forbidden in the Standard Model (SM) by lepton number (LN)
conservation. However, an extended version of the SM could contain
terms that violate LN and allow the $0\nu2\beta$ decay. Probable
mechanisms of LN violation may include exchanges by: Majorana
neutrinos $\nu_M$s, SUSY Majorana particles, scalar bilinears , e.g. doubly charged
Higgs, leptoquarks, right-handed $W_R$ bosons etc. \cite{moh}. These various contributions will have to
be disentangled to extract information from the $0\nu2\beta$ decay
on the characteristics of the sources of LN violation, in
particular, on the neutrino masses and mixing. Measurements of the
$0\nu2\beta$ decay in different nuclei will help to determine the
underlying physics mechanism~\cite{Deppisch:2006hb,Gehman:2007qg,Fogli}. In
Ref.~\cite{0706.4165} the $0\nu2\beta$ decay angular correlation for
the $^{76}$Ge nucleus was investigated in order to discriminate
among the various possible mechanisms contributing to this decay.
However much more new physics one can extract using the experimental
data for various decaying nuclei. In this report, we generalize the
analysis of Ref.~\cite{0706.4165} for the case of the following set
of nuclei: $^{76}$Ge, $^{82}$Se, $^{100}$Mo, $^{130}$Te, and
$^{136}$Xe.

~~~{\bf 2.} Following Ref.~\cite{0706.4165}, we use the general
effective Lagrangian for the $0\nu2\beta$ decay
\vspace{-0.3cm}
\begin{equation}\label{L}
    {\cal L}=\frac{G_{\rm F}V_{ud}}{\sqrt{2}}[(U_{ei}+\epsilon^{
    V-A}_{V-A,i})j_{V-A}^{\mu
    i}J^+_{V-A,\mu}+\sum\limits_{\alpha,\beta}\!^{^\prime}
\epsilon^\beta_{\alpha i}j^i_\beta J^+_\alpha+{\rm H.c.}]~,
\end{equation}
where the hadronic and leptonic currents are defined as:
$J^+_\alpha=\bar{u} O_\alpha d$ and $j^i_\beta =\bar{e} O_\beta
\nu_i$; the leptonic currents contain neutrino mass eigenstates
and the index $i$ runs over the light eigenstates; a summation
over the repeated indices is assumed;
$\alpha$,\,$\beta$=$V\!\mp\!A$,\,$S\!\mp\!P$,\,$T_{L,R}$
($O_{T_\rho}=2\sigma^{\mu\nu}P_\rho$,
$\sigma^{\mu\nu}=\frac{i}{2}\left[\gamma^\mu,\gamma^\nu\right]$,
$P_\rho=(1\mp \gamma_5)/2$ is the projector, $\rho=L,\,R$); the
prime indicates the summation over all the Lorentz invariant
contributions, except for $\alpha=\beta=V-A$, $U_{ei}$ is the PMNS
mixing matrix~\cite{PMNS} and $V_{ud}$ is the CKM matrix element
\cite{PDG}. The coefficients $\epsilon_{\alpha i}^\beta$ encode
new physics, parametrizing deviations of the Lagrangian from the
standard $V-A$ current-current form and mixing of the non-SM
neutrinos. Eq.~(\ref{L}) describes the so-called long range mechanism of the
$0\nu2\beta$ decay mediated by light Majorana neutrinos.

The differential width for the $0^+\!(A,Z)\rightarrow\!0^+\!(A,Z+2)
e^- e^-$ transitions is~\cite{0706.4165}
\begin{eqnarray}\label{dG}
d\Gamma / d\cos\theta = (\ln2/2)|M_{\rm GT}|^2{\cal
A}(1-K\cos\theta),\quad  K={\cal B}/{\cal A}~,\, -1<K< 1,
\end{eqnarray}
where $\theta$ is the angle  between the electron momenta in the
rest frame of the parent nucleus, $M_{\rm GT}$ is the Gamow--Teller
nuclear matrix element, and $K$ is the angular correlation coefficient.
Eq.~(\ref{dG}) is derived taking into account the leading
contribution of the parameters $ \epsilon_\alpha^\beta = U_{ei}\epsilon_{\alpha i}^\beta$.
The expressions for $\cal{A}$ and $\cal{B}$ for different choices of
$\epsilon_\alpha^\beta$, with only one nonzero coefficient considered  at a
time, are given in Ref.~\cite{0706.4165}.

Using the data on various decaying nuclei we have
considered the two particular cases for the parameter
space: A)~$\epsilon_\alpha^\beta=0$, $|\langle m \rangle|\neq0$~
(SM~plus Majorana neutrinos),  B)~$ \epsilon_\alpha^\beta\neq0,\ |\langle m \rangle|=0$
(vanishing effective Majorana mass). Only the terms with $\epsilon_{V\mp
A}^{V\mp A}$  are taken into account as the corresponding nuclear matrix elements have been
worked out in the literature \cite{Pantis}.

The differences in the half-lives and angular coefficients for
various  nuclei are described by the ratios ${\cal R}_\alpha^\beta(^A{\rm X})=T_{1/2}(\epsilon_\alpha^
 \beta,^A{\rm X})/T_{1/2}(\epsilon_\alpha^\beta,^{76}{\rm Ge})$ and ${\cal K}_\alpha^\beta(^A{\rm X})
 =K(\epsilon_\alpha^    \beta,^A{\rm X})/K(\epsilon_\alpha^\beta,^{76}{\rm Ge})$,
for the choice of only one nonzero coefficient $\epsilon_\alpha^\beta$
which characterize specific alternative new physics contributions (we make a comparison with $^{76}$Ge as it is the best tested isotope to date).
The numerical values of  $\cal{R}$, $\cal{K}$ and  $\cal{R}_\alpha^\beta$ , $\cal{K}_\alpha^\beta$ corresponding to cases A) and B), respectively,  are given in Tables~1 and 2 for two different nuclear models: QRPA without and with p-n pairing~\cite{Pantis}.
\vspace{-0.4cm}
\begin{table}[htb!]
\scriptsize
\caption{ The ratios of the half-lives $\cal{R}$ and
$\cal{R}_\alpha^\beta$ for various nuclei in QRPA without (with) p-n
pairing~\cite{Pantis}.}
\begin{center}
\begin{tabular}{|c||c|c|c|c|}
  \hline
Nucleus& $\cal{R}$=${\cal R}_{V-A}^{V-A}$\ &  ${\cal
R}_{V+A}^{V-A}$
   &  ${\cal R}_{V-A}^{V+A}$& ${\cal R}_{V+A}^{V+A}$\\
  \hline
  \hline
  $^{82}$Se & 0.42 (0.15) &  0.37 (2.76) &2.10 (3.07) &0.24 (0.03) \\
  \hline
  $^{100}$Mo &1.08 (195.18) &52.87 (0.59) & 1.11 (0.49) & 1.06 (0.79) \\
  \hline
  $^{130}$Te & 0.24 (0.11) & 0.21 (0.16) & 0.20 (0.12) & 0.15 (0.03) \\
  \hline
  $^{136}$Xe & 0.53 (0.15) & 0.40 (0.37) & 0.41 (0.22) & 0.34 (0.06) \\
  \hline
\end{tabular}
\end{center}
\end{table}
\begin{table}
\scriptsize
\caption{ The ratios of the angular coefficients $\cal{K}$ and
${\cal K}_\alpha^\beta$ for various nuclei in QRPA without (with)
p-n pairing~\cite{Pantis}.}
\begin{center}
\begin{tabular}{|c||c|c|c|}
  \hline
\ Nucleus\, & \ $\cal{K}$=${\cal K}_{V\pm A}^{V-A}$\,
   & \ ${\cal K}_{V-A}^{V+A}$\, & \ ${\cal K}_{V+A}^{V+A}$\\
  \hline
  \hline
  $^{82}$Se &  1.08 &  1.11 (1.11) & 1.13 (0.95)  \\
  \hline
  $^{100}$Mo & 1.08 & 1.14 (1.14) & 1.13 (0.84)  \\
  \hline
  $^{130}$Te & 1.04 & 1.07 (1.07)  & 1.01 (0.90)  \\
  \hline
  $^{136}$Xe & 1.03 & 1.06 (1.06)  & 0.98 (0.91) \\
  \hline
\end{tabular}
\end{center}
\vspace{-0.4cm}
\end{table}

The entries for the ratios of the
half-lifes are well separated, besides the ${\cal R}_{V-A}^{V-A}$, which is equal to $\cal{R}$. However, they are dominated by the
uncertainties of the nuclear model. On the other hand, the angular coefficients $\mathcal{K}$ and
${\mathcal{K}}_{V\pm A}^{V-A}$ do not depend on the nuclear matrix elements, and coefficients ${\mathcal{K}}_{V- A}^{V+A}$ essentially
do not depend on the uncertainties of the nuclear model. Moreover, the ratios of the angular correlations are not discriminating among the
various underlying theories as within the anticipated experimental uncertainty they are all consistent with unity. The most sensitive to the listed ratios
is $^{100}$Mo, except for the ratio ${\cal R}_{V-A}^{V+A}$ to which the most sensitive is $^{82}$Se. From the measurements of the half-lives, the most sensitive to the effects of $\epsilon_{V+A}^{V\pm A}$ and $\epsilon_{V-A}^{V+A}$ are the pairs $^{100}$Mo -- $^{130}$Te and $^{82}$Se -- $^{130}$Te, correspondingly. From the measurements of the angular coefficients,
the most sensitive to the effects of $\epsilon_{V\pm A}^{V+A}$ is the pair $^{76}$Ge -- $^{100}$Mo.

~~~{\bf 3.} In conclusion, the comparison of the half-lives and the electron angular correlations
for the selected decaying nuclei would help to minimize the theoretical uncertainties in the nuclear
matrix elements and identify the dominant mechanism underlying these
decays. At present, no experiment is geared to measuring the angular
correlations in $0\nu2\beta$ decays, as the main experimental thrust
is on establishing a nonzero signal unambiguously in the first
place. The running experiment NEMO3 has already
measured the electron angular distributions for the two neutrino
double beta decays of $^{100}$Mo and $^{82}$Se, and is capable of
measuring these correlations in the future for the $0\nu2\beta$
decays as well, assuming that the experimental sensitivity is
sufficiently good to establish these decays~\cite{NEMO3}. The
proposed experimental facilities that can measure the electron
angular correlations in the $0\nu2\beta$ decays are
SuperNEMO~\cite{SuperNEMO}, MOON~\cite{MOON}, and
EXO~\cite{EXO}.

\end{document}